\documentclass{ws-procs9x6}
\def\bfmat #1{\mbox{{\boldmath $#1$}}}
\begin{document}

\title{THE D'ALEMBERT-LAGRANGE PRINCIPLE FOR GRADIENT THEORIES AND BOUNDARY CONDITIONS}

\author{H. GOUIN}

\address{
Universit{\'e} d'Aix-Marseille, 13397 Marseille Cedex 20, France \\
E-mail: henri.gouin@univ-cezanne.fr\\ \vskip 0.30cm   {{ Dedicated
to Prof.\, Antonio M. Greco}}}

\begin{abstract}
Motions of continuous media presenting
singularities are associated with phenomena involving shocks,
interfaces or material surfaces. The equations representing
evolutions of these media are irregular
through   geometrical manifolds. \\
\noindent A unique continuous medium is conceptually simpler than
several media with surfaces of singularity. To avoid the surfaces of
discontinuity in the theory, we transform the model by considering a
continuous medium taking into account more complete internal
energies expressed in gradient developments associated with the
variables of state. Nevertheless, resulting equations of motion are
of an higher order than those of the classical models: they lead to
non-linear models associated with more complex integration processes
on the mathematical level as well as on the numerical point of view.
In fact, such models allow a precise study of singular zones when
they have a non negligible physical thickness. This is typically the
case for capillarity phenomena in fluids or mixtures of fluids in
which interfacial zones are transition layers between phases or
layers between fluids and solid walls. Within the framework of
mechanics for continuous media, we propose to deal with the
functional point of view considering globally the equations of the
media as well as the boundary conditions  associated with these
equations. For this aim, we revisit the {\it d'Alembert-Lagrange
principle of virtual works} which is able to consider the
expressions of the works of forces applied to a continuous medium as
a linear functional value on a space of test functions in the form
of {\it virtual displacements}. At the end, we analyze examples
corresponding to capillary fluids. This analysis brings us to
numerical or asymptotic methods avoiding the difficulties due to
singularities in simpler -but with singularities- models.
\end{abstract}

\section{Introduction}

A mechanical problem is generally studied through force interactions
between masses located in material points: this Newton point of view
leads together to the statistical mechanics but also to the
continuum mechanics. The statistical mechanics is mostly precise but
is in fact too detailed and in many cases  huge calculations crop
up. The continuum mechanics is an asymptotic notion coming from
short range interactions between molecules. It follows a loose of
information but a more efficient and directly computable theory. In
the simplest case of continuum mechanics, residual information comes
through stress tensor like Cauchy tensor
\cite{Germain1,Truesdell}\,. The concept of stress tensor is so
frequently used that it has become as natural as the notion of
force. Nevertheless, tensor of contact couples can be investigated
as in Cosserat medium \cite{Cosserat} or configuration forces like
in Gurtin approach \cite{Gurtin} with edge interactions of Noll and
Virga \cite{Noll}\,. Stress tensors and contact forces are
interrelated notions \cite{Isola2}\,.

\noindent A fundamental point of view in continuum mechanics is:\,
the Newton system for forces is equivalent to  \textit{the work of
forces is the value of a linear functional of  displacements. }Such
a method due to Lagrange is dual of the system of forces due to
Newton \cite{Germain2,Germain3} and is not issued from a variational
approach;  the minimization of the energy coincides with the
functional approach in a special variational principle only for some
equilibrium cases. \newline The linear functional expressing the
work of forces is related to the theory of distributions; a
decomposition theorem associated with displacements (as test
functions whose supports are $C^{\infty }\ $compact manifolds)
uniquely determines a canonical zero order form {\it (separated
form)} with respect both to the test functions and the transverse
derivatives of contact test functions \cite {Schwartz}\,.\newline
 As Newton's principle is useless when we do not
have any constitutive equation for the expression of forces, the
linear functional method is useless when we do not have any
constitutive assumption for the virtual work functional. The choice
of the simple material theory associated with the Cauchy stress
tensor corresponds with a constitutive assumption on its virtual
work functional. It is  important to notice that constitutive
equations for the free energy $\chi $ and constitutive assumption
for the virtual work
functional may be incompatible \cite{Gouin1}\,: for any \textit{%
virtual} displacement $\bfmat{\zeta}$ of an isothermal medium, the
variation $-\delta \chi \ $  must be equal to the \textit{virtual}
work of internal forces $\delta \tau _{int}$. The equilibrium state
is then obtained by the existence of a solution minimizing the free
energy. \newline
The equation of motion of a continuous medium is deduced from the \textit{%
d'Alembert-Lagrange principle of virtual works} which is an
extension of the principle in mechanics of systems with a finite
number of degrees of freedom: \textit{The motion is such that for
any virtual displacement the virtual work of forces is equal to the
virtual work of mass accelerations}.

\noindent Let us note: if the virtual work of forces is expressed in
classical notations in the form
\begin{equation}
\delta \tau =\int\int\int_{D}\left\{
\mathbf{f}.\,\bfmat{\zeta}+tr\left[ \left( -p\ {\mathbf 1} +2\mu \
\nabla \mathbf{V}\right). \nabla \bfmat{\zeta}\right] \right\}
dv+\int\int_{S }\mathbf{T}.\,\bfmat{\zeta}\ d{s } \label{viscous
fluid}
\end{equation}
from the  d'Alembert-Lagrange principle, we obtain not only the
equations of balance momentum for a viscous fluid in the domain $D$
but also the boundary conditions on the border $S$ of $D$. We notice
that expression (\ref{viscous fluid}) is not the Frechet derivative
of any functional expression.\newline If the free energy  depends on
the strain tensor $F$, then $\delta \tau $ must depend on $\nabla
\bfmat{\zeta}$ and leads to the existence of the Cauchy stress
tensor. If the free energy  depends on the strain tensor $F$ and on
the overstrain tensor $\nabla F,$ then $\delta \tau $ must depend
on $\nabla \bfmat{\zeta} $ and $\nabla ^{2}\bfmat{\zeta} $.\\

\noindent {\it Conjugated (or  transposed}) mappings being denoted
by asterisk, for any vectors ${\mathbf a}, {\mathbf b}$, we write
${\mathbf a}^*\, {\mathbf b}$\  for their {\it scalar product} (the
line vector is multiplied by the
  column vector) and
 ${\mathbf a} {\mathbf b}^*$  or $ {\mathbf a}\otimes {\mathbf b}$ for their
   {\it  tensor  product }
 (the column vector is multiplied by the line vector).
The product of a mapping   $\mathbf A$   by a vector   ${\mathbf a}$
is denoted by   $\mathbf A\,  {\mathbf a} $. Notation   ${\mathbf
b}^* \, \mathbf A\, $ means the covector   ${\mathbf c}^* $ defined
by the rule ${\mathbf c}^* = ( \mathbf A^*\,  {\mathbf b})^*$. The
divergence of a linear transformation   $\mathbf A$   is the
covector $ {\rm div} \mathbf A  $ such that, for any constant vector
${\mathbf a}, $ $ ({\rm div}\, \mathbf  A)\, {\mathbf a}    =   {\rm
div }\, (\mathbf A\ {\mathbf a}).\\
 $ We introduce a Galilean or fixed
system of coordinates $(x^{1},x^{2},x^{3})$ which is also denoted by
$\mathbf{x}$ as Euler or spatial variables. If $f$ is a real
function of $\mathbf x$, $ \displaystyle \frac {\partial f}{\partial
{\mathbf x}}$  is the linear form associated with the gradient of
$f$ and $\displaystyle \frac{\partial f} {\partial x^i}= (\frac
{\partial f} {\partial {\mathbf x}})_i $\,; consequently, $
\displaystyle (\frac {\partial f } {\partial {\mathbf x}})^*    =
{\rm grad} \, f$.
 The identity tensor is denoted by $\mathbf 1$.\\
Now, we present the method and its consequences in different cases
of gradient theory. As examples, we revisit the case of Laplace
theory of capillarity and the case of van der Waals fluids.
\section{Virtual work  of
continuous medium}
 The motion of a continuous medium is classically
represented by a continuous transformation $\mathbf{\varphi }$ of a
three-dimensional space into the physical space. In order to
describe the transformation analytically, the variables
$\mathbf{X=}(X^{1},X^{2},X^{3})$ which single out individual
particles correspond to material or Lagrange variables. Then, the
transformation representing the motion of a continuous medium is
\begin{equation*}
\mathbf{x=\varphi }\left( \mathbf{X,}t\right) \text{ \ or \
}x^{i}=\varphi ^{i}(X^{1},X^{2},X^{3},t)\, , \  i = 1, 2, 3
\label{motion}
\end{equation*}
where $t$ denotes the time. At $t$ fixed the transformation
possesses an inverse and continuous derivatives up to the second
order except at singular surfaces, curves or points. Then,
the\textbf{\ }diffeomorphism $\mathbf{\varphi }$ from the set
$D_{0}$ of the particles into the physical space $D$ is an element
of a functional space $ \mathbf \wp$ of the positions of the
continuous medium considered as a manifold with an infinite number
of dimensions.\newline
 To formulate
the  d'Alembert-Lagrange principle of virtual works, we introduce
the notion of {\it virtual displacements}. This is obtained by
letting the displacements arise from variations in the paths of the
particles. Let a one-parameter family of varied paths or
\emph{virtual motions} denoted by $\{\mathbf{\varphi }_{\eta }\}$
and possessing continuous derivatives up to the second order and
expressed analytically by the transformation
\begin{equation*}
\mathbf{x=\Phi }\left( \mathbf{X,}t;\eta \right) \label{vitual
motion}
\end{equation*}
with $\eta \in O,$ where $O$ is an open real set containing $0$ and
such that $\mathbf{\Phi }\left( \mathbf{X,}t;0\right) =\mathbf{\varphi }%
\left( \mathbf{X,}t\right) $ or $\mathbf{\varphi
}_{0}=\mathbf{\varphi }$
(the real motion of the continuous medium is obtained when $\eta =0$%
). The derivation with respect to $\eta $ when $\eta =0$ is denoted
by $\delta $. Derivation $\delta $ is named   \textit{variation} and
the {\it virtual displacement} is the variation of the position of
the medium \cite{Serrin}\,.  The
 virtual displacement is a tangent vector to $\mathbf{\wp }$ in $\mathbf{%
\varphi }$ ($\delta \mathbf{\varphi }\in T_{\mathbf{\varphi }}(\mathbf{\wp }%
))$. \   In the physical space, the {\it virtual displacement}
$\delta \mathbf{\varphi }$ is determined by the variation of each
particle: the {\it virtual displacement} \textit {of the particle} $\mathbf{x}$
 is such that $\bfmat{\zeta}=\delta \mathbf{x}\ $ when $\ \delta \mathbf{X}=0$, $\delta
\eta =1$ at $\eta =0$;  we associate the field of tangent vectors to
$D$
\begin{equation*}
{\mathbf{x}}\in D\ \mathbf{\rightarrow \bfmat{\zeta} }=\mathbf{\psi (x)}\equiv \frac{%
\partial \mathbf{\Phi }}{\partial \eta }\left| _{\eta
=0}\right. \in T_{\mathbf{x}}(D)
\end{equation*}
where $T_{\mathbf{x}}(D)$ is the tangent vector bundle to $D$ at
$\mathbf{x}$.
\begin{figure}[h]
\begin{center}
\includegraphics[width=9cm]{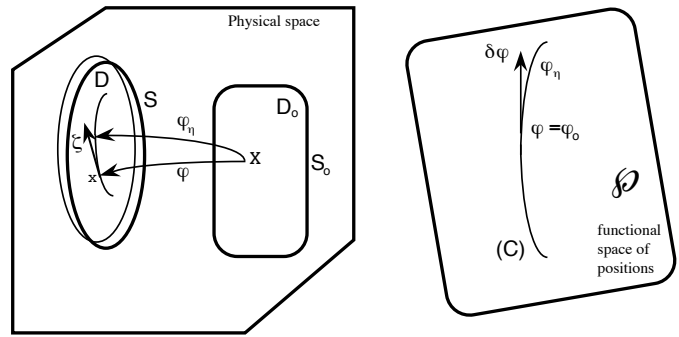}
\end{center}
\caption{The boundary $S$ of $D$ is represented by a thick curve and
its variation by a thin curve. Variation $\protect\delta
\mathbf{\protect\varphi}
$ of  family  $\{\mathbf{\varphi }_\eta \}$ of varied paths belongs to $T_{\mathbf{\protect\varphi }%
}(\mathbf{\wp })$, tangent space to $(\mathbf{\wp })$ at
$\mathbf{\varphi }$.} \label{fig1}
\end{figure}
The concept of virtual work  is purposed in the form:
\newline
\textit{The virtual work  is a linear functional value of the
virtual displacement,}
\begin{equation}
\delta \tau =<\Im ,\delta \mathbf{\varphi }>  \label{virtual work of
forces}
\end{equation}
where $< .\, , . >$ denotes the inner product of $\Im $ and $\delta
\mathbf{\varphi }$; then, $\Im $ belongs to the cotangent space of $\mathbf{%
\wp }$ at $\mathbf{\varphi }$ ($\Im \in \mathit{T}_{\mathbf{\varphi
}}^{\ast }(\mathbf{\wp }))$.\newline In Relation (\ref{virtual work
of forces}),  the medium in position $\mathbf{\varphi }$ is
submitted to the covector $\Im $ denoting all the stresses; in the
case of motion, we must add the inertial forces associated with the
acceleration quantities to the volume forces.\\The
d'Alembert-Lagrange principle
of virtual works  is expressed as:\\
\textit{%
\centerline {\small  \textbf{For all virtual displacements, the
virtual work  is null}.}}\newline \noindent Consequently,
representation (\ref{virtual work of forces}) leads to:
\begin{equation*}
\forall \ \delta \mathbf{\varphi }\in \mathit{T}_{\mathbf{\varphi }}(\mathbf{%
\wp }),\ \delta \tau =0  \label{d'Alembert-Lagrange}
\end{equation*}
\textbf{Theorem:} \textit{If expression (\ref{virtual work of
forces}) is a distribution in a separated form, the
d'Alembert-Lagrange principle yields the equations of motions and
boundary conditions in the form}\ \  $ \Im =0  $\,.

\section{Some examples of linear functional of forces}
Among all possible choices of linear functional of virtual
displacements, we classify the following ones:
\subsection{Model of zero gradient}
\subsubsection{Model A.0}
The medium fills an open set $D$ of the physical space and the
linear functional is in the form
\begin{equation*}
\delta \tau =\int \int \int_{D}F_{i}\, \zeta^{i}dv \label{grad A0}
\end{equation*}
where$\;F_{i} \, (i = 1,2,3) $ denote the covariant components of
the volume force $\mathbf{F}$ (including the inertial force terms)
presented as a covector. The equation of the motion is
\begin{equation}
\forall \;\mathbf{x}\in D,\ \,F_{i}=0
\;\Leftrightarrow\mathbf{\;F}=0 \label{gradient A0}
\end{equation}
\subsubsection{Model B.0}
The medium fills a set $D$ and the surface $S$ is the boundary of
$D$  belonging to the medium; with the same notations as in section
\textit{3.1.1}, the linear functional is  in the form
\begin{equation}
\delta \tau =\int \int \int_{D}F_{i}\,\zeta ^{i}dv + \int \int_S
T_{i}\,\zeta ^{i}d{s}  \label{grad B0}
\end{equation}
$T_{i}$ are the components of the surface forces (tension)
$\mathbf{T}$. From Eq. (\ref {grad B0}), we obtain the equation of
motion  as in Eq. (\ref{gradient A0}) and the boundary condition,
\begin{equation*}
\forall \;{\mathbf x}\in S,\  \,T_{i}=0\ \Leftrightarrow
\mathbf{\;T}=0
\end{equation*},
\subsection{Model of first gradient}
\subsubsection{Model A.1}
With the previous notations, the linear functional is in the form
\begin{equation*}
\delta \tau =\int \int \int_{D}\left( F_{i}\,\zeta ^{i}-\sigma
_{i}^{j}\,\zeta _{,j}^{i}\right) dv   \label{grad A1}
\end{equation*}
where $\sigma _{i}^{j}\,  (i,j = 1,2,3) $ are the components of the stress tensor $%
\mathbf{\sigma }\mathbf{.}$ Stokes formula gets back to the model
$\it B.0$ in the separated form
\begin{equation*}
\delta \tau =\int \int \int_{D}\left( F_{i}+\sigma _{i,j}^{j}\right)
\zeta ^{i}dv - \int \int_{S}n_{j}\sigma _{i}^{j}\ \zeta ^{i}d{s}
\end{equation*}
where  $n_{j}  \, (j=1,2,3) $ are the components of a covector which
is the annulator of the vectors belonging to the tangent plane at
the boundary $S$. It is not necessary to have a metric in the
physical space; nevertheless, for the sake of simplicity it is
convenient to use the Euclidian metric; the vector $\mathbf{n}$ of
components $ n^{j}  \, (j=1,2,3)$ represents the external normal to
$S$ relatively to $D$;  the covector $\mathbf{n}^\star\ $ is
associated with the components $n_j$. We deduce the equation of
motion
\begin{equation}
\forall \;\mathbf{x}\in D,\, \  \ F_{i}+\sigma _{i,j}^{j}=0  \
\Leftrightarrow\ \mathbf{F}+ \mathrm{div}\,\mathbf{\sigma} =
0\label{gradient A1}
\end{equation}
and the boundary condition
\begin{equation*}
\forall \;\mathbf{x} \in S,\, \   \ n_{j}\,\sigma _{i}^{j}=0\
\Leftrightarrow\ \mathbf{n}^\star\, \mathbf{\sigma} = 0
\end{equation*}
\subsubsection{Model B.1/0: (Mixed model with first gradient in $D$ and zero gradient on $S$)}
The linear functional is expressed in the form
\begin{equation*}
\delta \tau =\int \int \int_{D}\left( F_{i}\,\zeta ^{i}-\sigma
_{i}^{j}\,\zeta _{,j}^{i}\right) dv + \int \int_{S }T_{i}\,\zeta
^{i}\,ds   \label{grad B1/0}
\end{equation*}
Stokes formula yields the separated form
\begin{equation*}
\delta \tau =\int \int \int_{D}\left( F_{i}+\sigma _{i,j}^{j}\right)
\zeta ^{i}\,dv +\int \int_{S}\left( T_{i}-n_{j}\sigma
_{i}^{j}\right) \zeta ^{i}\,ds \hskip 2cm ({S.\emph{0}})
\end{equation*}
and we deduce the equation of motion in the same form as Eq. (
\ref{gradient A1}) and the boundary condition
\begin{equation*}
\forall \;\mathbf{x} \in S,\,  \ \ n_{j}\,\sigma _{i}^{j}=T_i  \
\Leftrightarrow\ \mathbf{n}^\star\, \mathbf{\sigma} = \mathbf{T}
\end{equation*}
Model $\it B.1/0$  is  the classical theory for elastic media and
fluids in continuum mechanics.
\subsubsection{Model B.1}
The linear functional is expressed in the form
\begin{equation}
\delta \tau =\int \int \int_{D}\left( F_{i}\,\zeta ^{i}-\sigma
_{i}^{j}\,\zeta _{,j}^{i}\right) dv + \int \int_{S }\left(
T_{i}\,\zeta ^{i}+\gamma _{i}^{j}\,\zeta _{,j}^{i}\right) ds
\label{grad  B1}
\end{equation}
where the tensor $\mathbf{\gamma }$ of components $\gamma _{i}^{j}$
is a new term.  The boundary of $D$ is a surface $S$ shared in a
partition of $N$
parts $S_{p}$ of class $C^{2}$, $(p=1,...,N)$ (Fig. 2). We denote by $%
(R_{m})^{-1}$ the mean curvature of $S$; the edge  $\Gamma _{p}$ of
$ S_{p}$ is the union of the limit edges $\Gamma _{pq}$ between
surfaces $S_{p}$ and $S_{q}$  assumed to be of class $C^{2}$ and
$\mathbf{t}$ is the tangent vector to $\Gamma_{p}$ oriented by
$\mathbf{n}$; $\mathbf{n}^\prime$
is the unit external normal vector to $\Gamma _{p}$ in the tangent plane to $%
S _{p}$: $\mathbf{n}^{\prime }\mathbf{=t\times n.}$   Let us notice
that:
\begin{equation}
\gamma _{i}^{j}\,\zeta _{,j}^{i}=-\gamma _{i,j}^{j}\,\zeta
^{i}+V_{,j}^{j} \label{int by parts}
\end{equation}
\begin{figure}[h]
\begin{center}
\includegraphics[width=9cm]{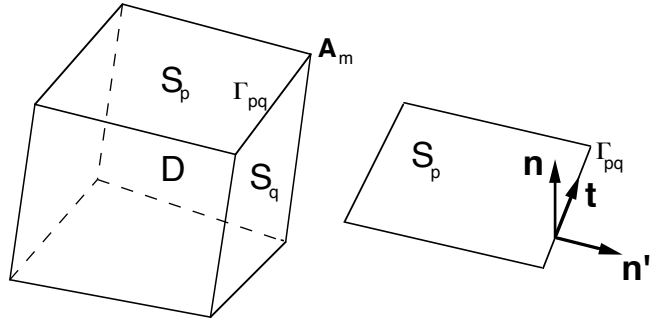}
\end{center}
\caption{The set $D$ has a surface boundary $S$ divided in several
parts. The edge of $S $ is denoted by $\Gamma $ which is also
divided in several parts with end points $\mathbf A_{m}$. }
\label{fig3}
\end{figure}
where $V^{j}=\gamma _{i}^{j}\,\zeta ^{i}\ $; consequently, from
integration of the divergence of vector $\mathbf V$ on surfaces
$S_p$ we obtain,
\begin{equation}
\int \int_{S _{p}}V_{,j}^{j}\;ds = -\int \int_{S _{p}}n_{j}\left(
\frac{V^{j}}{R_{m}}-V_{,l}^{j}n^{l}\right) ds + \int_{\Gamma
_{p}}n_{j}^{^{\prime }}V^{j}\,d\ell \label{int surface}
\end{equation}
We emphasize with the fact that $V_{,l}^{j}\,n^{l}$ corresponds to
the normal derivative to $S _{p}$  denoted $\dfrac{dV^{j}}{d{n}}$.
An integration by parts of the term $\sigma _{i}^{j}\,\zeta
_{,j}^{i}$ in relation (\ref {grad B1}) and taking account of
relations (\ref{int by parts}-\ref{int surface}) implies
$$
\delta \tau =\int \int \int_{D}F_{i}^{1} \zeta ^{i}\;dv + \int
\int_{S }T_{i}^{1} \zeta ^{i}\;ds+\int \int_{S }L_{i}\frac{d\zeta
^{i}}{dn} \, ds +\sum_{p=1}^{N}\int_{\Gamma _{p}}R_{pi}\,  \zeta
^{i}d\ell  \eqno{(S.\it 1)}
$$
with the following definitions
\begin{equation*}
\left\{
\begin{array}{ll}
\ F_{i}^{1}  \equiv F_{i}+\sigma _{i,j}^{j}\,,\ & L_{i}\equiv n_{j\;}\gamma _{i}^{j}\,  \\
\ T_{i}^{1}\equiv T_{i}-n_{j}\left(\sigma_i^j-\dfrac{d}{dn}(\gamma
_{i}^{j})+ \dfrac{1}{R_{m}}\gamma _{i}^{j}\right)-\gamma_{i,j}^j\,,
\; \ & R_{pi}\equiv n_{j}^{^{\prime }}\;\gamma _{i}^{j}\,
\end{array}
\right.
\end{equation*}
Due to theorem  {{\small {37}}} in \cite{Schwartz} , the
distribution (S.{\it 1}) has a unique decomposition in displacements
and transverse derivatives of displacements on the manifolds
associated with D and its boundaries:  expression $(S.{\it 1})$ is
in a separated form. Consequently, the equation of motion is
\begin{equation*}
 \forall \;\mathbf{x}\in D,\ F_{i}^{1}=0
\Leftrightarrow \    \mathbf{F}^1 = 0
\end{equation*}
and the boundary conditions are
\begin{equation*}
\begin{array}{ll}
\forall \;\mathbf{x}\in S, \  T_{i}^{1}=0,  L_{i}=0
\;&\Leftrightarrow\
 \ \mathbf{T}^1 = 0,\ \mathbf{L} = 0 \ \ \ \\
\forall \;\mathbf{x}\in \Gamma_{pq} , \ {R}_{pi}{+R}_{qi}=0\
 &\Leftrightarrow\  \ \mathbf{R}_p + \mathbf{R}_q = 0\
\end{array}
\end{equation*}
Term $\mathbf{L} $ is not reducible to a force: its virtual work
$L_{i}\dfrac{ d\zeta ^{i}}{dn}$ is not the product of a force with
the displacement $\bfmat{\zeta} $; the term $\mathbf{L} $  is an
{\it embedding action}.
\subsection{Model of second gradient}
\subsubsection{Model A.2}
The linear functional  is in the form
\begin{equation*}
\delta \tau =\int \int \int_{D}\left( F_{i}\,\zeta ^{i}-\sigma
_{i}^{j}\,\zeta _{,j}^{i}+S_{i}^{jk}\zeta _{,jk}^{i}\right) dv
\end{equation*}
Tensor $\mathbf{S}$ with $S_{i}^{jk}=S_{i}^{kj}$ is  an
\textit{overstress tensor}. An integration by parts of the last term
brings back to the model $\it B.1$,
\begin{equation*}
\delta \tau =\int \int \int_{D}\left( F_{i}\,\zeta ^{i}-\left(
\sigma _{i}^{j}+S_{i,k}^{jk}\right) \,\zeta _{,j}^{i}\right) dv +
\int \int_{S }n_{k}S_{i}^{jk}\ \zeta _{,j\ }^{i}\, ds
\end{equation*}
and the virtual work  gets the separated form $({S.\it 1})$ with:
\begin{equation*}
\left\{
\begin{array}{cc}
F_{i}^{1}=F_{i}+\sigma _{i,j}^{j}+S_{i,jk}^{jk} & \text{volume force} \\
T_{i}^{1}=-n_{j}\left( \sigma
_{i}^{j}+S_{i,k}^{jk}-\dfrac{d}{dn}\left( n_{k}S_{i}^{jk}\right)
+\dfrac{1}{R_{m}}n_{k}S_{i}^{jk}\right)  & \text{
surface force} \\
R_{pi}=n_{j}^{\prime }\ n_{k}\ S_{i}^{jk} & \text{ line force} \\
L_{i}=n_{j}\ n_{k}\ S_{i}^{jk} & \text{embedding action}
\end{array}
\right.
\end{equation*}
and consequently yields the same equation of motion and boundary
conditions as in case $\it B.1$.
\subsubsection{Model B.2}
The linear functional is in the form
\begin{equation*}
\delta \tau =\int \int \int_{D}\left(F_{i} \zeta ^{i}-\sigma
_{i}^{j} \zeta _{,j}^{i}+S_{i}^{jk}\zeta _{,jk}^{i}\right) dv +\int
\int_{S }\left( T_{i} \zeta ^{i}+\gamma _{i}^{j} \zeta
_{,j}^{i}+U_{i}^{jk}\zeta _{,jk}^{i}\right)ds \label{grad B2}
\end{equation*}
This functional yields two integrations successively on $S _{p}$ and
on $\Gamma _{pq}$ with terms at the points $A_{m}$. With obvious
notations, for the same reasons as in section {\it 3.2.3}, the
virtual work gets the separated form
\begin{eqnarray*}
\delta \tau &=&\int \int \int_{D}F_{i}^{1}\,\zeta ^{i}dv + \int
\int_{S }\left( T_{i}^{1}\,\zeta ^{i}+L_{i}^{1}\frac{d\zeta ^{i}}{dn}%
+L_{i}^{2}\frac{d^{2}\zeta ^{i}}{dn^{2}}\right) ds  \notag \\
&&{\small +}\sum_{p}\int_{\Gamma _{p}}\left( R_{pi}\ \zeta ^{i}+M_{pi}\frac{%
d\zeta ^{i}}{dn^{\prime }}\right) d\ell + \sum_{m}\phi _{mi}\,\zeta
_{\mathbf A_{m}}^{i}  \hskip 3cm ({S.\emph{2}})
\end{eqnarray*}
where $\zeta^i_{\mathbf A_{m}} \, (i =1,2,3)$ are the components of
$\bfmat{\zeta}$ at point $\mathbf A_m$.  The calculations are not
expended.\ They introduce the curvature tensor on $S _{p}$ and the
geodesic curvature of $\Gamma _{pq.}$. Consequently, $F_{i}^{1},\
T_{i}^{1},\,R_{pi},\ \phi _{mi}\ $are associated with volume,
surface, line and forces at points; $L_{i}^{1},\ L_{i}^{2},\ M_{pi}$
are embedding efforts of order 1 and 2 on $S$ and of order 1 on the
edge $\Gamma .\ $The equation of motion and boundary conditions
express that these seven tensorial quantities are null on their
domains of values  $D$, $S$, $\Gamma_p$ and $\mathbf A_m$.
\section{Conclusion}
It is possible to extend the previous presentation by means of more
complex medium with {\it gradient of order n}. The models introduce
embedding effects of more important order on surfaces, edges and
points. The {\it (A.n)} model refers to a {\it (B.n-1)} model: the
fact that boundary surface $S$ is (or is not) a material surface has
now a physical meaning. Consequently, we can resume the previous
presentation as follows:\newline  {\bf a})  The choice of a model
corresponds to specify the part ${\emph{G}}$ of the algebraic dual
$T_\varphi^*(\wp)$ in which the efforts   are considered: $\Im \in
{\emph{G}} \subset T_\varphi^*(\wp)$.
\newline {\bf b}) In order to operate with the  principle of virtual works
and to obtain the mechanical equations in the form $\Im  = 0$, it is
no matter that the part ${\emph{G}}$ of the dual is separating
$(\forall\, \Im \in {\emph{G}}, <\Im,\delta\varphi> =0\Rightarrow
\delta\varphi=0)$, but it is important the part $ {\emph{G}}$ is
separated $( \Im \in {\emph{G}}, \forall\, \delta \varphi \in
T_\varphi (\wp), <\Im,\delta\varphi> =0 \Rightarrow \Im = 0 )$.
\newline
{\bf c})   The functionals $\it A.1$, $\it B.1/0$, $\it A.2$, $\it
B.2$ are not separated: if $\Im$ consists in the data of the fields
$\mathbf F$, $\mathbf \sigma$, $\mathbf T$, it is not possible to
conclude that the fields are zero. \newline {\bf d}) Functionals in
$\it A.0$, $\it B.0$, $\it S.1$, $\it S.2\ldots $ are separated: if
the fields $\mathbf S^1$, $\mathbf T^1$, $\mathbf R^1$, $\mathbf L,
\ldots$ are continuous then, by using the fundamental lemma of
variation calculus, their values must be equal to zero. They are the
only functionals we must know  for using the principle of virtual
works; it is exactly as for a solid:   the torque of forces is only
known in the equations of motion.\newline {\bf e}) When the fields
are not continuous on surfaces or curves, we have to consider a
model of greater order in gradients and to introduce integrals on
inner boundaries of the medium.

For conservative  medium, the first gradient theory corresponds to
the compressible case.  The theory of fluid, elastic, viscous and
plastic media  refers to the model $\it (S.0)$. The Laplace theory
of capillarity in fluids  refers to the model $\it (S.1)$.
 To take into account superficial effects acting between solids and
fluids,
 we  use the model of fluids endowed with
capillarity ($\it S.2$);  the theory interprets the capillarity in a
continuous way and contains the Laplace theory of capillarity; for
solids,  the model corresponds to "elastic materials with couple
stresses" indicated by Toupin in \cite{Toupin}\,.
\section{Example 1: The Laplace theory of capillarity}
 Liquid-vapor and two-phase interfaces are
represented by a material surface endowed with an energy relating to
Laplace surface tension. The interface appears as a surface
separating two media with its own characteristic behavior and energy
properties \cite{Levitch}\, (when working far from critical
conditions, the capillary layer has a thickness equivalent to a few
molecular beams \cite {Ono}). The Laplace theory of capillarity
refers to the model $\it B.1$ in the form $(\it S.1)$ as following:
for a compressible fluid with a capillary effect on the wall
boundaries, the free energy  is in the form
\begin{equation*}
\chi =\int \int \int_{D}\rho \ \alpha (\rho )\ dv +
\sum_{p=1}^{N}a_{p}\int \int_{S_p }ds
\end{equation*}
where $\alpha(\rho)$ is the fluid specific energy, $\rho $ is the
matter density and  coefficients $a_{p}$ are the surface tensions of
each surface $S_p$.
 Surface integrations are associated to the space metric;
 the virtual work of internal forces is
\begin{equation*}
\delta \tau_{int} =\int \int \int_{D}-p_{,i}\, \zeta ^{i}dv
+\sum_{p=1}^{N}\int \int_{S _{p}}n_{i}\left( p-\dfrac{a_{p}}{R_{m}}%
\right) \,\zeta ^{i}\, ds + \sum_{p=1}^{N}\int_{\Gamma
_{p}}a_{p}n_{i}^{\prime }\, \zeta ^{i}\, d\ell
\end{equation*}
where $p\equiv \rho^2\alpha^\prime(\rho)$ is the fluid pressure. The
external force (including inertial forces) is the body force $\rho\,
\mathbf{f} $ defined  in $D$, the surface force  is $\mathbf{T}$
defined  on $S$ and the line force is $\mathbf{R}$ defined  on
$\Gamma$. D'Alembert-Lagrange principle  yields the equation of
motion and boundary conditions:
\begin{equation*}
\forall \;\mathbf{x}\in D,\  -p_{,i}+\rho \, f_{i}=0 \ \
\Leftrightarrow\ \  -{\rm grad}\, p + \rho\,  \mathbf{f} = 0,
\end{equation*}
\begin{equation*}
\begin{array}{cc}
\forall \;\mathbf{x}\in S,\  \ \ p\
n_{i}+T_{i}-a_p\dfrac{n_{i}}{R_{m}} = 0\  \ \Leftrightarrow\ \ p\,
\mathbf{n}+\mathbf{T}- a_p
\dfrac{\mathbf{n}}{R_m} = 0 \\
\forall \;\mathbf{x}\in \Gamma_{pq},\  \ a_{p}n_{pi}^{\prime
}+a_{q}n_{qi}^{\prime }+R_{i}=0          \ \Leftrightarrow\ \     \
a_{p}\mathbf{n}^\prime_p +a_{q}\mathbf{n}^\prime_q +\mathbf{R} = 0
\end{array}
\end{equation*}
Boundary conditions are   {\it Laplace equation}  and {\it
Young-Dupr\'{e} condition}.
\section{Example 2: Fluids endowed with internal
capillarity}

For interfacial layers,  kinetic theory of gas leads to laws of
state associated with non-convex internal
energies\cite{cahn,rowlinson}\,. This approach  dates back to van
der Waals \cite{vdW}\,, Korteweg \cite{Korteweg}\,, corresponds to
the Landau-Ginzburg theory \cite {Hohenberg} and presents two
disadvantages. First,  between phases, the pressure may become
negative; simple physical experiments can be used  to cause traction
that leads to these negative pressure values \cite{Rocard}\,.
Second, in the field between bulks, internal energy cannot be
represented by a convex surface associated with density and entropy;
this fact seems to contradict the existence of equilibrium states;
it is possible to eliminate this disadvantage by writing in an
anisotropic form the stress tensor of the capillary layer which
allows
to study  interfaces of non-molecular size near a critical point.\\
One of the problems that complicates this study of phase
transformation dynamics is the apparent contradiction between
Korteweg  classical stress theory and the Clausius-Duhem inequality
\cite{Gurtin2}\,. Proposal made by Eglit \cite{Eglit}\,, Dunn and
Serrin \cite{Dunn}\,, Casal and Gouin \cite{Gouin5} and others
rectifies this  anomaly for liquid-vapor interfaces. The simplest
model in continuum mechanics considers a free energy  as the sum of
two terms: a first one corresponding to a medium with a uniform
composition equal to the local one and a second one associated with
the non-uniformity of the fluid \cite{cahn,vdW}\,. The second term
is approximated by a gradient expansion, typically truncated to the
second order. The model is simpler than models associated with the
renormalization-group theory \cite {domb} but has the advantage of
easily extending well-known results for equilibrium cases to the
dynamics of interfaces \cite {slemrod,trusk,GouinRuggeri}\,.\\ We
consider a fluid $D$ in contact with a wall $S$.
 Physical experiments prove that
the fluid is nonhomogeneous in the neighborhood of $ S
 $ \  \cite{rowlinson}\,.  The internal
energy $ \varepsilon $ is also a function of the entropy. In the
case of isothermal motions, the  internal energy is replaced by the
 free energy. In the mechanical case, the entropy and the
temperature are not concerned by the virtual displacements of the
medium. Consequently, for  isentropic or isothermal motions,   $
\varepsilon  =
 f(\rho , \beta )
$ where  $ \beta  = ({\rm grad}\, \rho)^2 $.   The fluid is
submitted to external forces represented by a  potential
 $ \Omega $  as a function of Eulerian
variables $ {\mathbf x}$. To obtain boundary conditions  it is
necessary
 to know  the wall effect.
An explicit
 form for the energy of
interaction between surfaces and liquids is proposed in
\cite{Gouin3}\,.
 We denote  by $B$ the surface density of energy at the wall. The total energy $
E $ of the fluid  is the sum of three potential energies:
 $E_f$ (bulk energy), $E_p$ (external energy)  and $E_S$ (surface
 energy).
\begin{equation*}
E_f = \int\int\int_D\rho\, \varepsilon (\rho , \beta)\ dv  ,
 \  E_p\ =\
\int\int\int_D\rho\, \Omega({\mathbf x}) \ dv    , \  E_S\ = \
\int\int_S  B\ ds
\end{equation*}
We have the   results  (see   Appendix):\ \ $ \displaystyle\delta
E_f = \int\int\int_D(-{\rm div }\ {\sigma}) \, {\bfmat{\zeta}}\ dv$
\begin{equation*}  +  \int\int_S \left\{ -A {\frac  {d{\zeta}_n}{dn}}\ +
\left(\frac{2A}{R_m} \, {\mathbf n}^* + {\rm grad}_{tg}^*A+ {\mathbf
n}^*\sigma
 \right) {\bfmat{\zeta}}\right\} ds - \int_\Gamma A\, {\mathbf n^{\prime *}}\,
{\bfmat{\zeta}}\, d\ell
\end{equation*}
with  \ \ $\displaystyle \  \sigma=  - P {\mathbf 1} - C \ {\rm
grad}\; \rho \otimes {\rm grad}\; \rho \equiv - P {\mathbf 1} - C\,
(\frac{\partial \rho}{\partial {\mathbf x}} )^* \, \frac{\partial
\rho}{\partial {\mathbf x}}$, where $ C =
2\rho\,\varepsilon^\prime_{\beta}$, $P = \rho^2
\varepsilon^\prime_\rho- \rho \   {\rm div} (C{\rm\ grad}\, \rho) $,
$ \varepsilon^\prime_\rho $ (or $ \varepsilon^\prime_\beta $)
denoting the partial derivative of $\varepsilon$  with respect to $
\rho $ (or $\beta$), ${\zeta}_n={\mathbf n}^*\,{\bfmat{\zeta}} $;
 $A=C\rho\ \displaystyle\frac{d\rho}{dn}\ $  where $\displaystyle\
\frac{d\rho}{dn}=\frac {\partial \rho}{\partial {\mathbf x}}\
{\mathbf n}$ and $  {\rm grad}_{tg} $ denotes the tangential part of
the gradient relatively to $S$.
\begin{equation*}
\delta E_p=\int\int\int_D\ \rho\  \frac{\partial \Omega}{\partial
 {\mathbf x}} \, {\bfmat{\zeta}}\ dv \equiv \int\int\int_D\ \rho\
({\rm grad}^*\, \Omega) \,  {\bfmat{\zeta}}\ dv; \ \ {\rm and},
\end{equation*}
\begin{equation*}
\delta E_S=\int\int_S\left\{\delta B- \left({2B\over R_m}\,{\mathbf
n}^* + {\rm grad}^*_{tg}B \right)  {\bfmat{\zeta}} \right\}ds +
\int_\Gamma B\, {\mathbf n^{\prime *}}\, {\bfmat{\zeta}}\, d\ell
\end{equation*}
The density in the fluid has a limit value $\rho_{_S} $ at the wall
$S$ and  $ B $ is assumed to be a function of $ \rho_{_S} $ only
\cite{Gouin 3} . Then, $ \delta B = B^\prime(\rho_{_S})\,
\delta\rho_{_S} = -\rho_{_S} B^\prime({\rho_{_S}})\,{\rm div} \,
{\bfmat{\zeta}} $, where ${\rm div} \,  {\bfmat{\zeta}}$ is computed
on $S$\, \cite{Serrin} .
 Let us denote $ G =-\rho_{_S}  B^\prime(\rho_{_S})$; Appendix yields
\begin{equation*}
 \int\int_S \delta B\ ds =\int\int_S G\
{\rm div}\, {\bfmat{\zeta}}\ ds =
\end{equation*}
\begin{equation*}
=\int\int_S\ \left\{ G\, \frac{d \zeta_n}{dn} - \left (\frac
{2G}{R_m}\, {\mathbf n}^*  +{\rm grad}^*_{tg} G \right )
{\bfmat{\zeta}} \right\}  ds  + \int_\Gamma G\, {\mathbf n^{\prime
*}}\, {\bfmat{\zeta}}\, d\ell
\end{equation*}
\begin{equation*}
\delta E_S=\int\int_S\left\{ G\,\frac {d{\zeta}_n} {dn} -
\left(\frac{2H}{R_m} \,{\mathbf n}^*  +{\rm grad}^*_{tg}H \right)
{\bfmat{\zeta}}\right\} ds + \int_\Gamma H\, {\mathbf n^{\prime
*}}\, {\bfmat{\zeta}}\, d\ell
\end{equation*}
with $ H = B(\rho_{_S})-\rho_{_S} B^\prime  (\rho_{_S})$. Then,
\begin{equation}
\begin{array}{ll}
 \delta E
 = \displaystyle\int\int\int_D(\rho\ {\rm grad}^*\, \Omega -{\rm
div}\, \sigma)\, {\bfmat{\zeta}}\, dv\  - \int_\Gamma (A-H)\,
{\mathbf n^{\prime *}}\, {\bfmat{\zeta}}\, d\ell  \\
 + \displaystyle\int\int_S (G-A)  \frac {d{\zeta}_n}{dn} + \left (
\frac {2(A-H)}{R_m}\,{\mathbf n}^* + {\rm grad}^*_{tg}(A-H)
+{\mathbf n}^* \sigma \right)    {\bfmat{\zeta}}\ ds \label{S3}
\end{array}
\end{equation}
At equilibrium,  $ \delta\tau\equiv -\delta E=0 $. The fundamental
lemma of variation calculus associated with separated form
(\ref{S3}) corresponding to ($S.\emph{2}$), yields:

\subsection{Equation of equilibrium}

From any arbitrary variation  ${\mathbf x}\in D \rightarrow
{\bfmat{\zeta}} ({\mathbf x})\ $ such that ${\bfmat{\zeta}}
={\mathbf 0}$ on $
 S $, we get
\begin{equation*}
\int \int \int _D\left({\rho \ {\rm grad}^* \Omega   - {\rm div}
\,\sigma }\right){\bfmat{\zeta}}\, ds=0. \ \  {\rm Then},
\end{equation*}
\begin{equation}
 \rho \  {\rm grad}^* \Omega - {\rm div}\, \sigma =0
\label{motion5}
\end{equation}
This equation is written in the classical form of equation of
equilibrium \cite{Gouin5} . {\it It is not the same for the boundary
conditions.}
\subsection{Boundary conditions}
\subsubsection{Case of a rigid  wall}
We consider a rigid wall; on $S$, the virtual
 displacements satisfy  the
condition  $ {\mathbf n}^*  \, {\bfmat{\zeta}} =0$.
 Then, at the rigid wall, $\forall\  {\mathbf x} \in S \rightarrow {\bfmat{\zeta}}({\mathbf x})$ such that
 $ {\mathbf n}^*  \, {\bfmat{\zeta}}
 =0$,
\begin{equation*}
 \int\int_S (G-A)  \frac {d{ \zeta}_n}{dn} + \left (
\frac {2(A-H)}{R_m}\,{\mathbf n}^* + {\rm grad}^*_{tg}(A-H)
+{\mathbf n}^* \sigma \right)    {\bfmat{\zeta}}\ ds=0
\end{equation*}
Due to $\sigma=\sigma^*$, we deduce the boundary conditions
(\ref{conditions5}-\ref{conditions5.1})
\begin{equation}
\forall\  {\mathbf x}\in  S ,\
 G-A=0,\label{conditions5}
\end{equation}
and there exists a Lagrange multiplier  $ {\mathbf x} \in S
\rightarrow \lambda({\mathbf x}) \in R$ such that,
\begin{equation}
\forall\  {\mathbf x}\in  S ,\ \frac {2(A-H)}{R_m}\ {\mathbf n}  +
{\rm grad}_{tg}(A-H)+ \sigma\; {\mathbf n} =\lambda \, {\mathbf
n}\label{conditions5.1}
\end{equation}
The edge $\Gamma$ of $S$ belongs to the solid wall and consequently
on $\Gamma$, $ {\bfmat{\zeta}} = \eta\, {\mathbf t}$: the integral
on $\Gamma$ is null and does not yield any additive condition.
\subsubsection{Case of an elastic wall}
The equilibrium equation (\ref{motion5}) is unchanged. On $S,$ the
condition (\ref{conditions5})  is also unchanged. The only different
condition comes from the fact that we do not have anymore the
slipping condition for the virtual displacement on $S$,
  $({\mathbf
n}^*\,{\bfmat{\zeta}}=0)$. Due to the possible deformation of the
wall, the virtual work of
 stresses on $
S $ is $\displaystyle \delta E_e=\int\int_S{\bfmat
\kappa}^*{\bfmat{\zeta}}\ ds + \int_\Gamma {\mathbf
R}^*{\bfmat{\zeta}}\, d\ell $ where  $ {\bfmat \kappa}  = {\mathbf
\sigma_e}\, {\mathbf n} $ is the stress (loading) vector associated
with stress tensor  $\sigma_e$ of the elastic wall  and ${\mathbf
R}$ is the line force due to the elasticity of the line. Relation
(\ref{conditions5.1}) is replaced by
\begin{equation*}
 \forall\  {\mathbf x}\in  S ,\ 2\, \frac  {(A-H)}{R_m}\;{\mathbf n} + {\rm grad}_{tg}(A-H)+
\sigma\; {\mathbf n} + {\bfmat \kappa} = 0
\end{equation*}
We obtain an additive condition on $\Gamma$ in the form
$(H-A)\,{\mathbf n} ^\prime + {\mathbf R} = 0$ and due to condition
(\ref{conditions5}),
\begin{equation}
\forall\  {\mathbf x}\in  \Gamma ,\ B\,{\mathbf n} ^\prime +
{\mathbf
 R} = 0 \label{condition6}
\end{equation}
(If $\Gamma$ is the union of edges $\Gamma_p$, $ \ B\,{\mathbf n}
^\prime $ is replaced by $\displaystyle\sum_p\ B_p\,{\mathbf n}
^\prime_p$).

\subsection{Analysis of the boundary conditions}
Eq. (\ref{conditions5}) yields \ $
  \displaystyle C\ \frac {d\rho}{dn}+B^\prime (\rho_{_S})=0$;
 the definition of $\sigma$  implies:
\begin{equation*}
 \sigma \,{\mathbf n} = - P{\mathbf n} - C\; \frac {d\rho}{dn}\  {\rm grad}\;
 \rho
 \end{equation*}
Due to the fact that the tangential part of Eq.
(\ref{conditions5.1}) is always verified, the only condition comes
from Eq. (\ref{conditions5}); Eq. (\ref{conditions5.1}) yields the
value of the Lagrange multiplier $ \lambda$ and Eq.
(\ref{condition6}) the value of ${\mathbf
 R}$.  For an
elastic (non-rigid)
 wall we obtain,
\begin{equation}
\kappa_{tg}=0\ \   \hbox{ and }\ \ \kappa_n\ =P+\frac {2B}
 {R_m}-B^\prime(\rho_{_S})\,
\frac {d\rho}{dn}\label{conditions5.3}
 \end{equation}
where $ \kappa_{tg} $  and $ \kappa_n $ are the tangential and the
normal components of ${\bfmat \kappa}$. Taking into account of Eq.
(\ref{conditions5.3})  we obtain the stress values at the non rigid
elastic wall.  The surface energy is \cite{Gouin 3}\,:
$\displaystyle B(\rho_{_S})=-\gamma _1\,\rho_{_S} +\frac {\gamma
_2}{2}\,\rho_{_S}^2  $ where $  \gamma_1$ and  $ \gamma_2 $ are two
positive constants and the fluid density condition at the wall is
\begin{equation*}
C\, \frac {d\rho}{dn}=\gamma _1-\gamma _2\,\rho_{_S}
\end{equation*}
If we denote by $\displaystyle \rho_{_B}={\gamma _1}/{\gamma _2}$
the {\it bifurcation fluid density} at the wall, due to the fact $C$
is positive constant  \cite{rowlinson}\,, we obtain:\ if $ \rho_{_S}
< \rho_{_B} $,\ (\,or $\rho_{_S} > \rho_{_B} $), $\displaystyle
\frac {d\rho } {dn} $ is positive (\,or negative) and we have a lack
(\,or excess) of fluid density at the wall. Such media allow to
study fluid interfaces and interfacial layers between fluids and
solids and lead to numerical and asymptotic methods \cite{Gavri}\,.
\newline The extension to the dynamic
case is straightforward:  Eq. (\ref{motion5}) yields
\begin{equation*}
\rho\, {\mathbf  \Gamma}^* - {\rm div}\, \sigma+\rho\  {\rm
grad}^*\Omega =0
\end{equation*}
Vector ${\mathbf \Gamma}$ is the acceleration; boundary conditions
(\ref{conditions5}-\ref{conditions5.3}) are unchanged.

\section*{Acknowledgments}
I am grateful to Professor Tommaso Ruggeri for helpful discussions.

\section*{Appendix}

Let $\ S $ be a surface in the 3-dimensional space and $ {\mathbf n}
$ its external normal extended locally in the vicinity of $S$ by the
expression ${\mathbf {n(x)}} = {\rm grad}\  d(\mathbf {x}), $ where
$d$ is the distance of a point $\mathbf x$ to $S$; for any vector
field  $ {\mathbf w}$, we obtain \cite{Koba}\,:
\begin{equation*}
{\rm rot} ({\mathbf n} \times {\mathbf w}) ={\mathbf n }\, {\rm div}
\, {\mathbf w}-  {\mathbf w}\,{\rm div}\,{\mathbf n } + \frac
{\partial
  {\mathbf n}} {\partial {\mathbf x}}\,
 {\mathbf w}- \frac
{\partial
  {\mathbf w}} {\partial {\mathbf x}}\,
 {\mathbf n}
\end{equation*}
From $\,\displaystyle {\mathbf n}^*\frac {\partial
  {\mathbf n}} {\partial {\mathbf x}}= 0\,$ and $\,{\rm div}\,{\mathbf n }
  =\displaystyle -\frac
{2} {R_m}$ we deduce on $S$,
\begin{equation}
 {\mathbf n^*}  {\rm rot} ({\mathbf n} \times {\mathbf w}) = {\rm div} \,
  {\mathbf w}+\frac
{2} {R_m}\,  {\mathbf n^*}  {\mathbf w} - {\mathbf n^*} \frac
{\partial
  {\mathbf w}} {\partial {\mathbf x}}\,
 {\mathbf n}  \label{A0}
\end{equation}
We deduce: for any scalar field $ A $ and $  {\mathbf w} = A\,
{\bfmat{\zeta}}$,
\begin{equation}
 A\,\displaystyle {\rm div}\, {\bfmat{\zeta}}=A\, \frac{d{\zeta}_n} {dn}-\frac
{2A}{R_m}\,{\zeta}_n-({\rm grad}^*_{tg}A)   \, {\bfmat{\zeta}}+
{\mathbf n}^*  {\rm rot}\, (A{\mathbf n} \times{\bfmat{\zeta}})
\label{A1}
\end{equation}
 \textbf{Let us calculate}  $\delta E_f$; $D$ is a material
volume, then $\displaystyle \delta E_f= \int\int\int_D\rho\
\delta\varepsilon\,  dv$  with $\displaystyle\ \delta\varepsilon=
\frac {\partial \varepsilon }{\partial \rho}\ \delta\rho+ \frac
{\partial \varepsilon}{\partial \beta}\ \delta\beta$. From
$\displaystyle \delta\,\frac {\partial \rho}{\partial {\mathbf x}}=
\frac {\partial \delta\rho}{\partial {\mathbf x}}-\frac{\partial
\rho} {\partial {\mathbf x}}\, \frac {\partial
{\bfmat{\zeta}}}{\partial {\mathbf x}}$\ \ (see\ \ \cite{Gouin5}),
\begin{equation*}
\rho\,\varepsilon^\prime_\beta\, \delta\beta
=2\rho\,\varepsilon^\prime_\beta\, \delta  \frac {\partial
\rho}{\partial {\mathbf x}} \, \left(\frac {\partial \rho}{\partial
{\mathbf x}}\right)^*
 = C \left(\frac {\partial
\delta\rho}  {\partial {\mathbf x}}-\frac{\partial \rho} {\partial
{\mathbf x}}\,\frac {\partial {\bfmat{\zeta}}} {\partial {\mathbf
x}}\right)\left(\frac {\partial \rho}{\partial {\mathbf x}}\right)^*
\end{equation*}
\begin{equation*}
= {\rm div} (C\ {\rm grad}\, \rho\ \delta\rho) - {\rm div}(C\ {\rm
grad}\, \rho)\, \delta\rho-tr\left( C\, {\rm grad}\, \rho  \ {\rm
grad}^*\rho\ \frac {\partial {\bfmat{\zeta}}} {\partial {\mathbf
x}}\right)
\end{equation*}
Due to    $  \delta\,\rho=-\rho \,{\rm div}\, {\bfmat{\zeta}}$\ \
(see\ \ \cite{Serrin}),
\begin{equation*}
\begin{array}{cc}
\displaystyle \delta E_f=\int\int\int_D\left(\frac
 {\partial P}{\partial {\mathbf
x}}+{\rm div}(C\ {\rm grad} \ \rho\ {\rm grad}^*
\rho)\right){\bfmat{\zeta}}\, dv\\
\displaystyle\ -\int\int\int_D {\rm div} \left(C\,\rho\, {\rm
grad}\, \rho\  {\rm div} \, {\bfmat{\zeta}}+C\, {\rm grad} \, \rho\
{\rm grad}^*\,\rho \ {\bfmat{\zeta}}+P\,{\bfmat{\zeta}}\right) dv\\
 \displaystyle=\int\int\int_D -({\rm div}\,\sigma)\,{\bfmat{\zeta}}\,
dv+\int\int_S(-A\ {\rm div}\, {\bfmat{\zeta}}+{\mathbf n}^*\sigma \,
{\bfmat{\zeta}}) ds
\end{array}
\end{equation*}
From Eq. (\ref{A1}), we deduce immediatly: \ \ $\displaystyle \delta
E_f= \int\int\int_D(-{\rm div}\, \sigma){\bfmat{\zeta}}\ dv\ +$
\begin{equation*}
\displaystyle\int\int_S \left\{-A\,  \frac {d\zeta_n} {dn}+
\left(\frac {2A}{R_m}\ {\mathbf n}^* + {\rm grad}^*_{tg}\,A+{\mathbf
n}^*\sigma \right) {\bfmat{\zeta}}\right \}
 ds -\int_\Gamma A\,{\mathbf n'}^*\,{\bfmat{\zeta}}\, d\ell
\end{equation*}
\textbf{ Let us calculate} $\delta E_S$; due to  $\displaystyle\
E_S=\int\int_S B\ \det\,
 ({\mathbf n},d_1{\mathbf x},d_2{\mathbf x})\ $ where
  $\displaystyle\ d_1{\mathbf x} \ $ and
$\displaystyle\ d_2{\mathbf x}\ $ are two coordinate lines of $ S, $
we get:
\begin{equation*}
 E_S=\int\int_{S_0}B\, \det\, F\   \hbox{det}\, (F^{-1}{\mathbf n},d_1{\mathbf
X},d_2{\mathbf X})
\end{equation*}
 where $S_0 $ is the image of $S$ in a reference space with Lagrangian coordinates  ${\mathbf X}$ and
$ F  $ is the deformation gradient tensor $\displaystyle \frac
{\partial {\mathbf x}}{\partial {\mathbf X}}$ of components $\left
\{ \displaystyle \frac {\partial {x^i}}{\partial {X^j}}\right\}$.
Then,\ \ \ \ \ \ $\displaystyle \delta E_S=\int\int_{S_0}\delta B\
\det\, F\ \hbox{det}\, (F^{-1}{\mathbf n},d_1{\mathbf X},d_2{\mathbf
X})$
\begin{equation*} +
\int\int_{S_0}B\, \delta \left(\det\, F\ \hbox{det}\,
(F^{-1}{\mathbf n},d_1{\mathbf X},d_2{\mathbf X})\right ).
\end{equation*}
\begin{equation*}
\begin{array}{cc}
{\rm with} \ \ \ \displaystyle \int\int_{S_0}B\,  \delta
\left(\det\, F\ \hbox{det}\,
(F^{-1}{\mathbf n},d_1{\mathbf X},d_2{\mathbf X})\right) = \\
\displaystyle \int\int_S B\ {\rm div}\, {\bfmat{\zeta}} \ \det
 ({\mathbf n},d_1{\mathbf x},d_2{\mathbf x})  +  B\, \det
  \left (\displaystyle\
 \frac {\partial {\mathbf n}}{\partial {\mathbf x}}\,{\bfmat{\zeta}},d_1{\mathbf x},d_2{\mathbf x}\right )
\\
\displaystyle -  B  \det  \left (\displaystyle
 \frac {\partial{\bfmat{\zeta}}}{\partial {\mathbf x}}\,{\mathbf n},d_1{\mathbf x},d_2{\mathbf x}
  \right )= \int\int_S \left( {\rm div} (B\,{\bfmat{\zeta}}
)-({\rm grad}^* B) \,  {\bfmat{\zeta}}  -B {\mathbf n}^* \frac
{\partial {\bfmat{\zeta}}} {\partial {\mathbf x}} \, {\mathbf n}
\right ) ds
\end{array}
\end{equation*}
Relation (\ref{A0}) yields:  $\displaystyle {\rm div}\,
(B\,{\bfmat{\zeta}}) + \frac {2B} {R_m}\, {\mathbf n}^*
{\bfmat{\zeta}} - {\mathbf n}^* \frac {\partial B\, {\bfmat{\zeta}}}
{\partial {\mathbf x}} \, {\mathbf n}  = {\mathbf n}^*\, {\rm rot}\,
(B\,{\mathbf n}\times {\bfmat{\zeta}} )$,
\begin{equation*}
\begin{array}{cc}
\displaystyle \int\int_{S_0}B\,\delta   \left (\det\ F\,\hbox{det}\,
(F^{-1}{\mathbf n},d_1{\mathbf X},d_2{\mathbf X})\right ) =
\\
\displaystyle\ \int\int_{S_0} \left( - \frac {2B} {R_m} \, {\mathbf
n}^*+{\rm grad}^*B\,({\mathbf {nn}^*-\mathbf 1}) \right )
{\bfmat{\zeta}} \, ds+\int\int_S{\mathbf n}^*\ {\rm rot}\,
(B\,{\mathbf n}\times{\bfmat{\zeta}})\,ds
\end{array}
\end{equation*}
where $ {\rm grad}^*B\,({{\mathbf {nn}}^*-\mathbf 1})  $ belongs to
the
 cotangent plane to $S$; we obtain
\begin{equation*} \delta E_S= \int\int_S \left(
    \delta B  - \left( \frac  {2B}{R_m }\ {\mathbf
n}^*  +{\rm grad}_{tg}^*B  \right){\bfmat{\zeta}} \right) ds
+\int_\Gamma B\,{\mathbf n'}^*{\bfmat{\zeta}}\, d\ell .\label{A3}
\end{equation*}


\begin{thebibliography}{50}



\bibitem{Germain1} P. Germain, \emph{Cours de m\'{e}canique
des milieux continus}
 (Masson, Paris, 1973).

\bibitem{Truesdell} C. Truesdell, \emph{First course in rational continuum
mechanics} (Academic Press, New York, 1994).

\bibitem{Cosserat} E. Cosserat and F. Cosserat, \emph{Sur la th\'{e}orie des
corps d\'{e}formables}  (Hermann, Paris, 1909).

\bibitem{Gurtin} M.E. Gurtin, \emph{Configurational forces as basic concepts in
continuum physics}  (Springer, Berlin, 2000).

\bibitem{Noll} W. Noll  and E.G. Virga, \emph{{\ Arch. Rat. Mech. Anal.}}
 {\textbf{3}}, 1 {(1990)}.


\bibitem{Isola2} F. dell'Isola and P. Seppecher, \emph{Meccanica}
\textbf{32}, 33  (1997).


\bibitem{Germain2} P. Germain, \emph{ J. M\'{e}canique}, \textbf{12}, 235 (1973).

\bibitem{Germain3} P. Germain,  \emph{ S.I.A.M. \,J. Appl. Math.}
\textbf{25}, 556 (1973).

\bibitem{Schwartz}L.  Schwartz, \emph{Th\'{e}orie des
distributions}, Ch. 3  (Hermann, Paris, 1966).

\bibitem{Gouin1}  H. Gouin  and F. Gouin,  \emph{Mech. Res. Comm.}
\textbf{10}, 21 (1983).

\bibitem{Serrin} J. Serrin, Mathematical principles of classical fluid
mechanics, in  \emph{Encyclopedia of Physics VIII/1,} Ed: S.
Fl\"ugge, (Springer, Berlin, 1960) pp. 125-263.

\bibitem{Toupin} R.A. Toupin, \emph{Arch. Rat. mech. Anal.}
\textbf{11}, 385 (1962).

\bibitem{Levitch}  V. Levitch, \emph{Physicochemical Hydrodynamics}
(Prentice-Hall,  New Jersey, 1962).

\bibitem{Ono}  {S. Ono  and S. Kondo,} {Molecular theory of surface tension in
liquid}, in   \emph{Encyclopedia of Physics, X,} Ed: S. Fl\"{u}gge
 (Springer, Berlin, 1960).

\bibitem{cahn}  {J.W. Cahn  and  J.E. Hilliard,}   \emph{J. Chem. Phys.}
 \textbf{31}, 688  (1959).

\bibitem{rowlinson}  { J.S. Rowlinson. and B. Widom,} \emph{Molecular theory of
capillarity}  (Clarendon Press, Oxford, 1984).


\bibitem{vdW}  {J.D. van der Waals},
{\emph{Archives N\'{e}erlandaises}} {\textbf{28}}, 121
{(1894-1895)}.

\bibitem{Korteweg} J. Korteweg, {\emph{Archives N\'{e}erlandaises}}
 \textbf{{2,\ n}$^{o}\ ${6}}, 1  (1901).

 \bibitem{Hohenberg}  P.C. Hohenberg  and B.I. Halperin,
\emph{{Rev. Mod. Phys.}} \textbf{{49}}, 435 (1977).

\bibitem{Rocard} Y. Rocard, \emph{Thermodynamique}  (Masson, Paris, 1952).


\bibitem{Gurtin2} M.E.  Gurtin,   \emph{Arch.\ Rat. Mech.\ Anal.} \textbf{19},
339 (1965).

\bibitem{Eglit} M.E. Eglit,   \emph{J. Appl. Math. Mech.} \textbf{29},
351 (1965).

\bibitem{Dunn}   J.E. Dunn and J. Serrin,
 \emph{Arch.\ Rat. Mech.\ Anal.} \textbf{88}, 95 (1985).

\bibitem{Gouin5}  {P. Casal   and H. Gouin},
\emph{C. R. Acad. Sci. Paris}  \textbf{{300, II}}, 231 {(1985)}.

\bibitem{domb}  {C. Domb,}  \emph{The critical point}
(Taylor \& Francis,  London,  1996).

\bibitem{slemrod}  {M. Slemrod,} \emph{{\ Arch. Rat. Mech.
Anal.}} \textbf{{81}},  301  (1983).

\bibitem{trusk}  {L. Truskinovsky,}
{\emph{P.M.M.}} \textbf{{51}}, 777 {(1987)}.

\bibitem{GouinRuggeri} H. Gouin and T. Ruggeri,
\emph{ Eur. J. Mech B/ Fluids} \textbf{24}, 596 (2005).

\bibitem{Gouin3} H. Gouin, \emph{ J. Phys. Chem. B} \textbf{102},
1212
 (1998).


\bibitem{Gavri} H. Gouin and S.L. Gavrilyuk,   \emph{Physica A}, \textbf{268},
291 (1999).

\bibitem{Koba} S.  Kobayashi and K. Nomizu, \emph{Foundations of
differential geometry}, {vol. 1}  (Interscience Publ., New York,
1963).

\end{thebibliography}
\end{document}